# Un cadre de conception pour réunir les modèles d'interaction et l'ingénierie des interfaces

# – DRAFT –

Jérôme LARD, Frédéric LANDRAGIN, Olivier GRISVARD et David FAURE
CNRS – THALES

**Résumé :** Nous présentons HIC (conteneur d'interaction homme-système), un cadre général pour l'intégration de l'interaction avancée dans les cycles de développement logiciel. Nous montrons comment ce cadre permet de réconcilier les méthodes du génie logiciel (MDA, MDE) avec les modèles de conception tels MVC ou PAC. Nous illustrons notre approche grâce à deux implantations différentes pour deux contextes applicatifs différents : un patron de conception logicielle, MVIC® (modèle vue interaction contrôle), et un modèle d'architecture, MI (middleware d'interaction).

**Mots-clés :** Architecture d'interaction, modèles de conception, méthodes de développement, paradigme d'interaction, modèles d'interaction, ingénierie de l'interaction.

**Abstract:** We present HIC (Human-system Interaction Container), a general framework for the integration of advanced interaction in the software development process. We show how this framework allows to reconcile the software development methods (such MDA, MDE) with the architectural models of software design such as MVC or PAC. We illustrate our approach thanks to two different types of implementation for this concept in two different business areas: one software design pattern, MVIC® (Model View Interaction Control) and one architectural model, IM (Interaction Middleware).

**Keywords:** Interaction architecture, design models, development methods, interaction paradigm, interaction models, interaction engineering.

## 1. Introduction

Cette étude se situe dans le contexte du prototypage et de l'utilisation des interfaces homme-machine (IHM) dites tactiques pour les domaines professionnels tels que la gestion et le contrôle du trafic aérien, la surveillance maritime ou encore la supervision de situation sur un terrain d'opérations. La réflexion que nous avons menée sur les besoins en termes d'interaction avancée pour les applications fournies par THALES, nous a conduit à proposer des solutions pour intégrer de nouvelles modalités et techniques d'interaction. Nous avons cherché également à améliorer le confort d'utilisation des applications. Nous avons ainsi identifié deux problèmes majeurs auxquels nous avons souhaité apporter des éléments de réponse.

Premièrement, un écart existe entre, d'un côté les nombreuses théories proposant des modèles pour la réalisation de logiciels interactifs multimodaux et adaptatifs, et de l'autre côté leurs déploiements dans les contextes industriels. Certains modèles ont été testés et utilisés avec succès (Abed et al.,1998 ; Kolski et al. ; 2003 ; Moussa et al., 2005). Pour autant, les modèles de conception logicielle issus de la recherche restent peu employés et les systèmes multimodaux sont rarement déployés.

Deuxièmement, nous remarquons que les approches du génie logiciel telles que l'initiative MDA (Model Driven Architecture), pour la standardisation d'une approche à base de modèles dans l'industrie logicielle, et le processus outillé correspondant MDE (Model Driven Engineering), basé sur l'utilisation des modèles et de leur transformation, ne prennent pas suffisamment en compte la problématique de l'interaction homme-machine et les exigences des utilisateurs.

L'objectif de cet article est de présenter une approche basée sur la modélisation des interactions qui cherche à réconcilier les théories portant sur les modèles d'architecture logicielle et les réalisations industrielles. Cette approche doit fournir un support à l'intégration de la gestion de l'interaction homme-machine dans le cadre du développement de systèmes d'information complexes et professionnels. Nous présentons les processus de développement des logiciels dans des domaines industriels. Nous faisons ensuite un état des lieux des méthodes et des modèles proposés pour représenter l'interaction. Puis, nous revenons sur le concept de HIC (Human-system Interaction Container), développé dans le laboratoire LINC de THALES (Sedogbo et al., 2003) et qui propose un cadre conceptuel pour les réalisations concrètes que nous exposons ensuite. Ces réalisations sont d'une part, le modèle MVIC® (modèle vue interaction contrôle), et d'autre part, le middleware d'interaction.

Le modèle architectural MVIC® est présenté dans son implantation actuelle sur un cas applicatif du domaine naval. Nous montrons dans ce premier cas comment nous avons réussi à modéliser et à intégrer l'interaction pour le développement d'interfaces graphiques. Le middleware d'interaction permet d'accueillir des services dits d'interaction, dédiés à la gestion du dialogue homme-machine. Ce Middleware a fait l'objet d'une implantation sur le cas de la gestion du trafic aérien.

Cette vision architecturale de la gestion de l'interaction, qui s'inspire des architectures dites N-tiers, offre une séparation entre la logique de persistance, la logique d'application, la logique d'interaction et la logique de présentation des données. Elle permet une adaptation à différents terminaux et à différents contextes de travail. La généralisation de cette approche à une architecture pour les systèmes interactifs est une nouveauté et permet d'intégrer de nombreux modèles de conception logicielle et de communication homme-machine.

## 2. Modèles de conception d'interfaces et méthodes de génie logiciel

### 2.1. Contraintes et exigences du point de vue industriel

Les concepteurs des outils et des architectures pour le prototypage des IHM à THALES cherchent à utiliser les modèles architecturaux ou conceptuels issus des laboratoires de recherche qui prennent en compte les problématiques d'adaptabilité aux différents contextes et d'adaptation aux différents moyens d'interaction. Parmi ces modèles certains prennent en compte la variabilité des moyens d'accès (Dragicevic et al., 2004 ; Huot et al., 2004) ou les évolutions des contextes physiques du système : changement de terminal (Thévenin et al., 1999 ; Calvary et al., 2001), évolution dans les modalités employées (Bouchet et al., 2004 ; Rousseau et al., 2006).

Cependant, dans de nombreux cas, les modèles de conception logicielle sont développés pour des applications éloignées de celles que nous rencontrons. Les spécificités réelles de certaines situations opérationnelles que nous devons prendre en compte demandent une adaptation à des contextes variés pour des situations se révélant souvent critiques, complexes et urgentes à résoudre. Les IHM développées sont dépendantes de plusieurs paramètres : évolution des organisations humaines au cours du temps ; forte distribution et forte hétérogénéité des données à présenter ; différents types de capteurs ; et enfin, variabilité des besoins en interaction pour un opérateur selon son rôle et ses préférences dans l'organisation.

La première des exigences est de fournir des moyens par lesquels les concepteurs de systèmes peuvent réaliser et tester des prototypes en s'appuyant sur de nouveaux outils et de nouveaux modèles pour l'interaction. Les modèles utilisés doivent être intégrés au sein d'une plate-forme de développement reposant sur une architecture logicielle adaptée à la conception de systèmes interactifs. Cette plate-forme doit être assez générique pour proposer une aide à la conception et à l'intégration de modèles d'interaction quel que soit le domaine concerné (naval, aéronautique, etc.).

La seconde exigence est celle de l'adéquation des besoins des utilisateurs avec le modèle d'interaction choisi dans le développement du système d'information. Nous présentons des moyens pour intégrer ces exigences et nous proposons un outil pour le support au développement des IHM.

## 2.2. Modèles et méthodes de génie logiciel

A l'heure actuelle, de nombreuses entreprises du secteur industriel développant des produits logiciels souhaitent généraliser l'utilisation d'approches basées sur les modèles. Dans ce cadre, MDA est une initiative encadrée par l'OMG (Object Management Group), soutenue par des industriels, des éditeurs de logiciels, et aussi par les futurs utilisateurs des systèmes réalisés selon cette méthode. MDA propose de décrire les applications au niveau de deux modèles, appelés PIM (Platform Independant Model) et PSM (Platform Specific Model). Ces modèles proposent aux développeurs de penser leurs applications à un niveau d'abstraction plus élevé. Ceci permet de prendre du recul par rapport aux évolutions technologiques rapides des plates-formes d'exploitation tout en proposant d'être plus proche des exigences des clients lorsqu'il s'agit d'adapter les logiciels à de nouveaux besoins. En résumé, les PIM sont des modèles indépendants des plates-formes d'exécution et expriment les spécificités de l'organisation cliente, alors que les PSM permettent de spécifier les plates-formes d'exécution.

UML (Unified Modeling Language) (Rumbaugh et al., 1998) est le langage de choix pour l'écriture de ces modèles. Une fois ces modèles spécifiés, la génération du code spécifique aux organisations et aux plates-formes est réalisée par les outils conçus dans cette optique. MDE propose donc des outils et des méthodes pour appliquer de manière systématique, ces concepts aux développements logiciels, là où MDA propose de nouvelles architectures et des standards. Mais ce type d'approche s'intéresse particulièrement à la réalisation d'un système informatique et au développement de ses fonctions. MDE fait peu de place aux pratiques de prototypage qui nous renseignent sur les contextes d'utilisation des logiciels ainsi que sur leur adéquation aux attentes des utilisateurs finaux. L'intégration de ces connaissances dans les modèles reste faible. Or, l'utilisabilité du système est déterminante à son acceptation, et peut être améliorée grâce à l'utilisation de modèles d'interaction homme-machine en amont des phases de conception logicielle.

Face à ces approches, nous proposons donc que les IHM soient réalisées sous la forme de modèles informatiques qui répondent à cette exigence d'abstraction des organisations et des plates-formes cibles. Le fait de matérialiser l'interaction homme-machine comme composant logiciel dans un cas (MVIC®) et comme couche logicielle fournissant des services remplissant un contrat de service pour l'interaction homme-machine dans l'autre (middleware d'interaction), permet aux concepteurs de spécifier et de développer l'interaction en tant que telle. La question qui se pose est alors la suivante : comment pouvons-nous intégrer ces connaissances dans les processus de développement de l'industrie logicielle ?

**2.3. Modèles autour de l'interaction homme-machine : une classification**

Les modèles d'interaction offrent une description opérationnelle de la manière dont l'interaction est implantée. Un modèle d'interaction est un ensemble de principes, de règles et de propriétés qui guident la conception des interfaces (Beaudouin-Lafon, 2004). Le modèle d'interaction décrit comment combiner un certain nombre de techniques d'interaction entre elles de manière cohérente. Ce modèle possède des propriétés qui permettent d'évaluer des conceptions interactives. Par exemple, l'interaction instrumentale (Beaudouin-Lafon, 2000 ; 2004) étend la manipulation directe (Shneiderman, 1998) avec l'introduction d'un nouveau tiers entre l'utilisateur et le logiciel : l'instrument.

Les architectures d'interaction décrivent les éléments fonctionnels de l'application interactive et leurs relations respectives (rôle des composants). Nous pouvons citer l'exemple de la triade MVC (Krasner et al., 1988) constituée par une relation du patron Observateur/Observé entre les vues et le modèle, et la relation du patron Composite entre les vues elles-mêmes. L'architecture d'interaction est dédiée à l'implantation des interfaces. Les modèles d'architecture et de conception tels que PAC (Coutaz, 1987), PAC-Amodeus (Nigay, 1993) sont très connus mais encore peu utilisés dans le monde industriel. Ceci est encore plus vrai pour des modèles plus anecdotiques comme MVP (Potel, 1996). En comparaison, MVC semble être plus répandu dans le monde industriel sans doute à cause de son intégration au sein des langages de programmation tels que SmallTalk (Krasner et al., 1988) puis Java. Cette intégration a permis de proposer des boîtes à outils exploitant les propriétés de ce modèle sous-jacent.

Les techniques d'interaction représentent la palette des outils matérialisant la fonction de transfert d'une action d'un utilisateur vers une interaction avec les objets des applications. Nous pouvons citer par exemple la technique de Drag'n Drop (Wagner et al., 1995), ou le Pan&Zoom (Bederson et al., 1994 ; Furnas et al., 1995 ; Bourgeois et al., 2002), qui sont deux techniques d'interaction caractérisées et évaluées.

Les boîtes à outils intégrant l'interaction fournissent des interfaces de programmation aux développeurs de systèmes interactifs. Par exemple, ICON (Dragicevic, 2004), MaggLite (Huot et al., 2004), ou encore SwingState (Appert et al., 2006) fournissent des capacités techniques pour la mise en œuvre de techniques d'interaction.

Les ateliers de conception logicielle sont identiques dans leur utilisation aux boîtes à outils, mais ils intègrent souvent un modèle d'architecture en plus de proposer des techniques. Ils proposent également une plate-forme pour le support à l'exécution du code développé. Ils peuvent également émettre des recommandations basées sur le modèle qu'ils implantent, en rapport avec des standards de développement logiciel tels que nous les abordons dans la prochaine section.

Cette classification nous est très utile car elle fournit un cadre théorique intéressant pour le développement de logiciels interactifs. Le modèle MVIC® (un modèle d'architecture) est présenté comme une recommandation pour le développement de composants logiciels de l'atelier JAGUAR©, comme nous le verrons dans la suite.

## 3. HIC, conteneur d'interaction homme-machine

HIC propose un principe de séparation fonctionnelle entre les différentes logiques implantées dans les logiciels existants, séparation illustrée dans la figure 1. HIC généralise une vision de la conception de l'interaction aux systèmes complexes pour lesquels les interactions sont fortement distribuées et fortement contraintes.

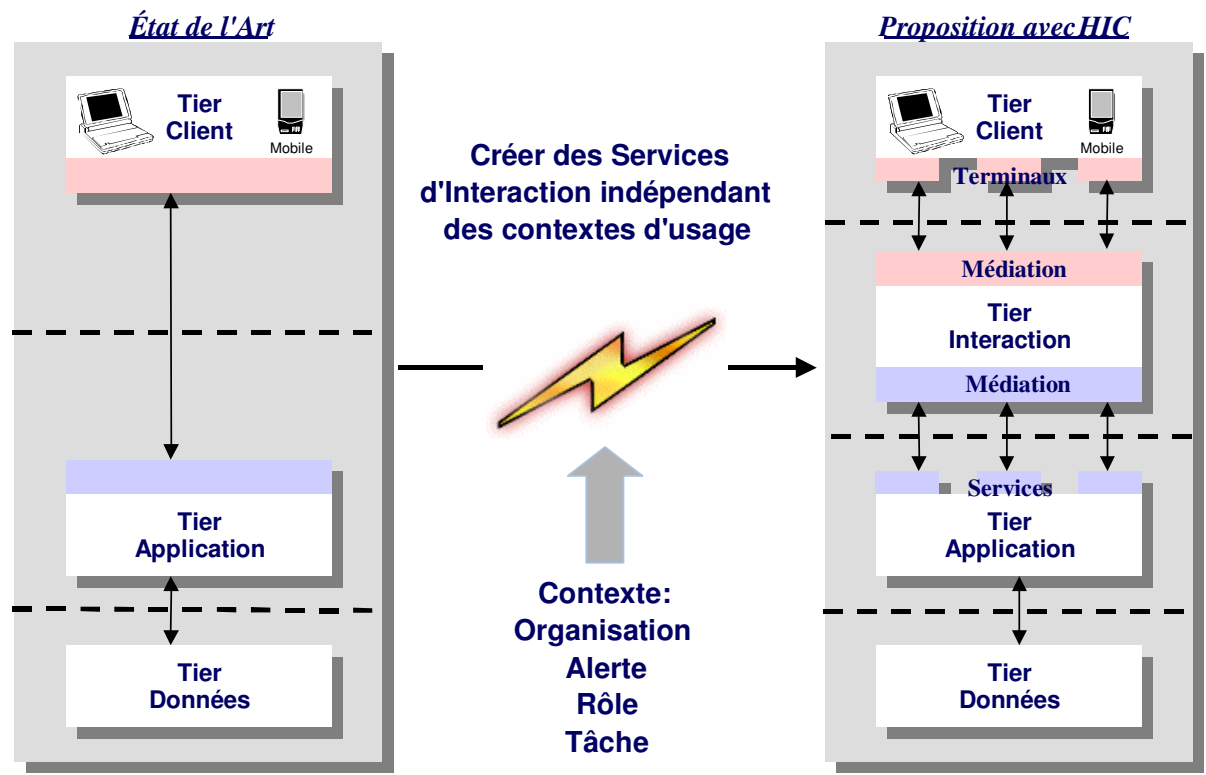

Figure 1. HIC et séparation des logiques dans un système N-tiers

L'introduction d'une gestion de l'interaction homme-machine dans ces systèmes entraîne une modification des architectures classiques dites 3-tiers, à trois niveaux d'abstraction (données, application, présentation) vers des architectures incluant une abstraction des interactions avec le ou les utilisateurs de l'application. Une nouvelle couche logicielle a été ajoutée qui représente dans l'architecture, les traitements informatiques dédiés à l'interaction homme-machine : la logique d'interaction. Un des moyens de proposer des interfaces hautement adaptatives à la tâche de l'opérateur, à son contexte ainsi qu'à son rôle, passe par l'introduction de ce tiers interaction séparant la logique de présentation de la logique métier. Il réalise une abstraction des terminaux et des modalités d'une part, et de l'application d'autre part. HIC est un cadre de conception qui fournit un modèle d'architecture fonctionnelle et propose d'encapsuler l'ensemble des composants logiciels dédiés à la gestion de l'interaction dans un conteneur sensible au contexte de son environnement d'exécution. De ce fait, ce conteneur agit comme un médiateur entre la logique applicative et la logique de présentation. HIC se rapproche donc des modèles d'architecture logicielle tels que ARCH (Bass et al., 1991) ou MVC (Krasner et al., 1988), du fait de la séparation des préoccupations entre application et interface, et la présence d'un contrôleur de dialogue, dans notre cas un médiateur, pour le contrôle de l'interaction entre l'utilisateur et le système.

Le découpage proposé par HIC a donc pour but de permettre de générer dynamiquement ou d'ajuster des interactions en fonction des besoins et des attentes des utilisateurs par rapport à leurs tâches et à leurs missions. Le contexte d'interaction sera évalué afin de définir ou de redéfinir la présentation des données mais aussi les capacités fonctionnelles offertes aux utilisateurs. L'interaction est adaptée par un changement dans l'enchaînement des actions fonctionnelles. Ces actions fonctionnelles sont celles qui sont relatives aux fonctions ou aux méthodes proposées par le système. En se basant sur HIC, nous proposons une gestion de l'interaction homme-machine orientée par le domaine et centrée sur les utilisateurs. Les deux implantations présentées illustrent les deux parties de notre approche. La première concerne l'abstraction des connaissances de l'interaction homme-machine au sein d'un modèle dédié à cet usage, et la seconde permet la projection de ces connaissances à l'exécution du système informatique dans une infrastructure logicielle, le middleware d'interaction.

## 4. MVIC®, modèle, vue, interaction, contrôle

### 4.1. Principes

De nombreux systèmes sont actuellement basés sur le patron de conception MVC (modèle, vue, contrôleur). Un patron de conception logiciel est une solution générale à un problème de conception récurrent. Ce patron de conception permet un développement simplifié d'une application. MVC permet de concevoir un logiciel interactif de manière robuste en séparant la partie de la logique applicative de la partie de la logique de présentation. Cependant, malgré la modularité de ce patron et la bonne séparation des responsabilités des composants de cette triade, il est difficile d'y définir la partie chargée de l'adaptation aux contextes opérationnels. Les contextes opérationnels sont constitués de multiples variables telles que l'organisation, les missions, les rôles, les tâches, etc. (cf. figure 1). En plus, les domaines d'applications cibles présentent à la fois des contraintes importantes pour la présentation de l'information mais également pour la définition des composants du domaine. Enfin, certaines technologies de programmation des interfaces homme-machine ont présumé que le composant de contrôle était fortement corrélé à la vue, ce qui a pour conséquence d'enfouir et de disperser ce contrôle au sein de l'application, et donc d'intégrer parfois la logique d'interaction dans la logique de présentation. Tel est le cas de la librairie Swing du langage de programmation Java qui encapsule l'apparence des contrôles (texture, couleur, taille, etc.) avec le code de gestion des événements utilisateurs sur ces contrôles (les écouteurs d'événements). Ceci a pour conséquence de n'offrir que des comportements interactifs standardisés souvent pauvres quant aux possibilités qui sont offertes aux développeurs. Les chercheurs et les ingénieurs se sont rendus compte de ces limitations et ont cherché à y remédier (Appert et al., 2006). La colocalisation du code relatif aux aspects des objets graphiques et à la logique de l'interaction dans les mêmes objets se révèle handicapante dans de nombreuses situations pour lesquelles l'interaction doit être adaptée en fonction du contexte opérationnel. L'approche « centrée utilisateur » implique la modification du patron de conception MVC tel qu'il a été fourni dans certaines boîtes à outils, vers un patron plus souple à interaction variable selon les contextes opérationnels.

### 4.2. Contexte applicatif

Nous proposons de formaliser l'interaction pour permettre une meilleure séparation des différents rôles de chacun des composants dans MVIC. Cette séparation et cette nouvelle identification des rôles des composants nous amène à une plus forte potentialité pour la réutilisation des différents modules. Les vues, qui, dans l'architecture classique MVC s'abonnaient au modèle pour présenter ses informations, s'abonnent désormais au module d'interaction quel que soit le rôle de l'opérateur.

Ainsi la réutilisation des vues sera plus grande. Cette idée étend la gestion des vues par le contrôleur à la présentation des objets d'intérêts du métier de l'opérateur. Selon les rôles des utilisateurs dans l'organisation, les besoins de représentation graphique s'exprimeront dans ce composant gérant l'interaction. Par exemple, une piste radar doit être un objet graphique métier configurable selon les besoins de l'utilisateur, dans son contexte opérationnel particulier. La présentation d'une piste à l'écran doit, à terme, devenir une fonction plus indépendante des modèles sous-jacents grâce aux abstractions mises en place.

L'externalisation des connaissances sur l'interaction dans des modèles spécifiques permet d'augmenter la réutilisation des objets reposant sur ces connaissances. Nous pouvons par exemple imaginer une nouvelle application identique à la précédente, pour laquelle un opérateur aurait des fonctions altérées ou augmentées dans un rôle particulier.

Grâce à cette nouvelle architecture, le module d'interaction peut disposer de connaissances sur le terminal de l'utilisateur (type de terminal, place restante sur son écran, nombre d'écrans). Ces informations peuvent lui permettre d'adapter la présentation à l'utilisateur en fonction du contexte (état de l'utilisateur, tâche en cours, rôle actuel de l'utilisateur). Il est possible d'utiliser la modélisation des tâches opérateurs (Faure et al., 2004) spécifique à chaque rôle particulier. Le changement de rôle de l'utilisateur signifie un changement de modèle de tâche. Les services applicatifs et les appels à la vue sont automatiquement altérés et réagissent dynamiquement aux nouvelles configurations. L'utilisation des connaissances sur l'état de l'utilisateur dans sa tâche permet de garantir la consistance de la mission à l'utilisateur et lui offre une gestion de son contexte d'interaction tout en conservant l'historique de ses actions avec le système. Ceci est utile en cas de panne d'une console par exemple.

Enfin, l'utilisation de ce patron permet une ouverture vers de nouvelles fonctionnalités qui facilitent le travail des concepteurs et fournissent une meilleure utilisabilité aux opérateurs. Nous pouvons citer à titre d'exemples : l'adaptation au terminal de l'utilisateur lui permettant de visualiser une situation altérée sur un terminal moins perfectionné, ou de disposer d'un résumé de l'affichage d'un écran dans une petite fenêtre ; l'utilisation de nouvelles modalités comme le geste ou la voix qui nécessitent une gestion plus fine du dialogue avec l'utilisateur ;

la génération automatique de fonctionnalités de présentation configurées grâce aux connaissances stockées dans les modèles pour les adapter au contexte (type de données à présenter, rôle actuel de l'utilisateur, progression dans sa tâche, etc.).

**4.3. Intégration d'un modèle d'interaction à la conception logicielle**

Ce premier cas montre de quelle manière nous avons conçu l'interaction plutôt que l'interface (Beaudouin-Lafon, 2004) et comment ce modèle d'interaction sert de base à une gestion de l'interaction plus avancée. Nous avons implémenté cette vision au sein de JAGUAR© (JAva Graphic Unified ARchitecture), un atelier de prototypage logiciel existant à THALES, fondé sur une approche à base de modèles de type MDE. JAGUAR© a été réalisé pour la spécification et le prototypage des IHM du domaine naval. Cet atelier de prototypage et de génération des IHM est basé sur le langage Java et sur le modèle d'architecture MVIC. JAGUAR© offre une modélisation complète des composants nécessaires au développement des applications sous la forme de modèles UML, ce qui permet une gestion efficace du développement des applications. L'atelier permet également la génération de documentation ainsi que la gestion des conflits de versions entre les différents modèles créés. Ce type de développement adopte les méthodes de travail proposées par les initiatives telles que MDA et illustre les procédés de fabrication logicielle dans l'industrie.

La figure 3 illustre l'architecture générale de l'atelier logiciel JAGUAR© et reprend donc de manière ad hoc les propositions présentes dans le modèle MVIC. Basé précédemment sur le modèle d'architecture logiciel MVC, l'atelier utilise désormais deux nouveaux composants afin de proposer une plus forte adaptativité aux conditions opérationnelles. Les composants DisplayModel et DisplayController constituent ainsi la partie consacrée à la gestion avancée de l'interaction. Le DisplayController en particulier joue ce rôle d'adaptateur pour l'interaction proposée à l'utilisateur en redéfinissant les actions du contrôleur de l'application (le composant Controller, ou le C de MVC).

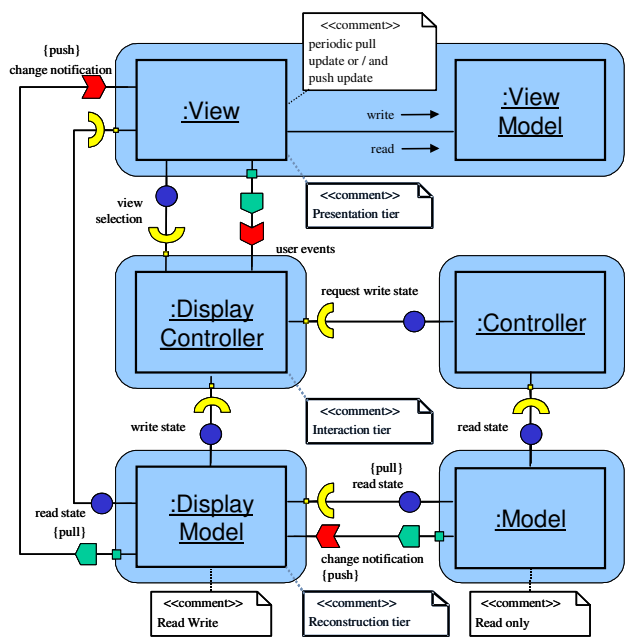

Figure 3. Le modèle d'architecture logicielle du framework JAGUAR©

Cet atelier conçu par des experts techniques du développement des IHM tactiques s'appuie sur un modèle de composants, appelé JACOMO (JAva Component Object MOdel), et qui a également été élaboré à THALES. Les composants JACOMO de l'architecture JAGUAR© représentés dans la figure 3 sont les suivants :
- le composant de présentation : View/ViewModel ;
- le composant d'interaction : DisplayController ;
- le composant de synthèse : DisplayModel ;
- le composant fonctionnel (métier) : ApplicationControl/Controller ;
- le composant d'application (données) : ApplicationModel/Model.

La vue (View) constitue une représentation des données de l'application (Model). Dans le cas de JAGUAR©, cette vue ne signifie pas automatiquement une représentation graphique des données car elle peut être adaptée à différents canaux de représentation (graphique, audio, texte, etc.). Une application se décompose en plusieurs modèles MVC dont chacun est responsable d'une sous-partie de l'interaction avec le client. Le modèle de la vue est spécifique à un type de vue particulier. Il conserve les données à afficher dans le type utilisé par la vue.

Le composant DisplayController constitue un cadre général de la gestion de l'interaction dans JAGUAR© et fournit dans cette architecture une gestion des interactions entre l'utilisateur et le système. Cette brique logicielle est partageable entre plusieurs vues au sein d'une même plate-forme. Ce composant est capable de gérer différents types de configuration pour l'interaction homme-machine. Le DisplayController constitue une passerelle qui renseigne quelles actions fonctionnelles doivent être appelées pour des événements produits par les utilisateurs. Le DisplayController est un contrôleur générique pour l'interaction et matérialise dans cet atelier le tiers médiateur entre l'utilisateur et le système. Le DisplayController représente le tiers interaction entre la vue et le contrôleur (entre la couche de présentation et la partie fonctionnelle de l'application).

Le composant DisplayModel est le modèle qui permet de reconstruire des données d'affichage, en particulier si l'IHM souhaite combiner plusieurs modèles sources pour constituer une donnée plus riche. Par exemple, ce composant peut associer ou agréger des données provenant de plusieurs autres composants Model, et construire une donnée à afficher qui a un intérêt particulier pour l'utilisateur alors qu'elle n'existait pas en tant que telle dans les données de l'application.

Le composant de contrôle, Controller, est un contrôleur classique : il gère le comportement de l'application. Il répercute les actions de DisplayController au Model, modifiant les données du système. Le composant Model enregistre et fournit les données de l'application au système. Il maintient l'état et les données du système. Lorsque des changements interviennent dans le modèle celui-ci en informe toutes ses vues. Ceci est matérialisé dans la figure 3 par l'événement « change notification ». Lorsque les vues abonnées à ce modèle sont notifiées elles peuvent récupérer les données correspondantes par un appel de service « read state ».

Ainsi la mise en place du tiers interaction identifié dans un composant logiciel, DisplayController, a été réalisée, ce qui a permis une modélisation des interactions pour les IHM du domaine naval. L'introduction de la gestion de l'interaction est une base technique à une gestion plus fine de la prise en compte des contextes opérationnels, en particulier de la gestion des alertes et de la distribution des rôles opérateurs. L'introduction de ce nouveau composant agissant comme médiateur entre l'utilisateur et le système, et jouant le rôle d'adaptateur du noyau fonctionnel de l'application, nous procure trois avantages majeurs : une plus grande réutilisation des autres composants, un accroissement de l'adaptabilité aux besoins opérationnels, et plus de possibilités dans l'interaction fournie.

## 5. MI, middleware d'interaction

### 5.1. Principes

La matérialisation et l'implantation que nous présentons dans cette section nous emmènent dans un domaine de recherche connexe, celui des middlewares. Au premier abord, un middleware est un composant logiciel d'interconnexion, qui consiste en un ensemble de services permettant à de multiples processus de s'exécuter sur une ou plusieurs machines et d'interagir à travers un réseau. Plus proche de notre point de vue, un middleware est aussi et surtout un composant logiciel qui constitue une couche de conversion et de traduction entre deux processus. Comment les principes de HIC peuvent-ils se matérialiser en un middleware, et pourquoi proposons-nous la notion de middleware d'interaction ? C'est à ces questions que nous répondons tout au long de cette section, avec un positionnement rapide par rapport aux principaux middlewares existants et une description de celui que nous avons conçu et implémenté. Nous complètons cette description par une illustration de l'opérationnalité de notre middleware d'interaction dans un contexte applicatif réel relevant de la gestion du trafic aérien.

Pour mieux appréhender cette notion de middleware, reprenons une classification classique des types de middleware (Issarny et al., 1998 ; Ranganathan et Campbell, 2003 ; Román et al., 2002 ; Schantz et al., 2002) :
- middlewares systèmes : il s'agit d'une couche de conversion insérée entre un flux de données brutes produit par une machine spécifique et le système d'exploitation de la machine ;
- middlewares explicites : il s'agit d'une couche intermédiaire entre le système d'exploitation et n'importe quelle application exécutée dessus. Ce type de middleware fournit une automatisation de quelques processus qui étaient auparavant inclus dans l'application elle-même (et souvent dans chaque application). Deux exemples de tels middlewares actuellement utilisés sont ceux de CORBA (COmmon Request Broker Architecture) et de RMI (Remote Method Invocation). Les middlewares explicites présentent parfois une organisation hiérarchique des services qu'ils fournissent en trois parties ou couches : la couche des services de distribution ; la couche des services communs ; la couche des services spécifiques au domaine (Schantz et al., 2002) ;
- middlewares implicites : il s'agit cette fois d'un processus de médiation ou d'interprétation entre une application métier et l'application de présentation qui lui est associée. Java offre un tel type de middleware, dans la mesure où les objets créés avec le langage peuvent être analysés et présentés en utilisant une API (Application Programming Interface) pour la représentation graphique : librairies javax.swing.* ou java.awt.*.

C'est dans la lignée de ces middlewares implicites que nous nous plaçons, avec notre découplage des logiques de présentation (IHM) et d'application métier. Mais la notion de middleware d'interaction que nous introduisons ne fait pas qu'insérer une couche de conversion entre ces deux processus de présentation et de métier. Nous lui ajoutons un rôle d'analyse et d'abstraction dirigé par les caractéristiques de l'interaction homme-machine. L'objectif est d'une part, de mieux comprendre les messages provenant de l'utilisateur, et en particulier d'identifier les intentions de celui-ci, et d'autre part, de mieux comprendre l'intérêt pour l'utilisateur d'une information provenant d'une application ou d'un capteur physique. Cette abstraction distingue notre approche de celle de certaines architectures comme les systèmes multi-agents (Martin et al., 1999 ; Tarpin-Bernard et al., 1999). Là où la plupart des architectures existantes ne proposent pas d'échanges pro-actifs et ont connaissance à l'avance, de manière le plus souvent statique, des producteurs et des consommateurs de messages typés, nous visons à donner aux infrastructures logicielles un rôle plus fonctionnel vis-à-vis des contextes d'utilisation et à identifier de manière dynamique sémantique des messages et identité des émetteurs et récepteurs pertinents.

Avant de décrire de manière précise les fonctionnalités et les composants du middleware d'interaction, il nous semble nécessaire de remettre en rapport cette matérialisation avec nos objectifs. Tout d'abord, en quoi cette nouvelle notion répond-elle au problème concernant l'écart existant entre d'une part, les modèles et les théories sur l'interaction homme-machine, et d'autre part, les réalisations industrielles ? Les réalisations industrielles que nous connaissons dans THALES font une part de plus en plus grande aux systèmes de systèmes ou d'une manière générale aux systèmes complexes. Notre approche consistant à implémenter dans une couche spécifique des composants précis qui sont essentiels à ces systèmes complexes va dans le sens de la généricité et la réutilisabilité, point de passage obligatoire pour des réalisations viables. Les prochaines tendances majeures de l'industrie logicielle sont prévues pour être orientées vers une approche de déploiement de services de haut niveau, et notre middleware d'interaction va dans ce sens. Ensuite, en quoi notre proposition répond-elle à une meilleure intégration de l'interaction dans les processus de développement ? Fournir une plate-forme générique de services va dans le sens de la facilitation des processus de développement logiciel industriel. En effet, nous pouvons imaginer qu'une entreprise récupère des composants d'autres entreprises pour son propre cas d'utilisation, surtout quand ces composants reposent sur une expertise complexe et pluridisciplinaire (on pensera par exemple à la commande vocale ou à la commande multimodale). L'expertise réside alors dans la chaîne complète, dans l'agencement des différents composants que le middleware d'interaction a contribué à clarifier.

### 5.2. Description des caractéristiques du middleware d'interaction

Le middleware d'interaction tel que nous l'avons conçu fournit d'une part, un noyau d'interaction capable de gérer des événements utilisateurs et de les mettre en rapport avec les spécificités d'une ou de plusieurs applications, et d'autre part, un ensemble de services d'interaction, c'est-à-dire un certain nombre d'outils relatifs à l'interaction homme-machine et utilisables par n'importe quel composant à l'intérieur ou à l'extérieur du middleware. Deux notions s'avèrent ainsi fondamentales : celle d'événement et celle de service.

Pour bâtir notre middleware d'interaction, il nous fallait partir d'une plate-forme capable de gérer des événements et des services. Compte tenu de ces prérequis, ainsi que des besoins supplémentaires tels que la gestion d'un réseau avec les caractéristiques de répartition et de distribution des ressources associées, notre choix s'est porté sur la plate-forme iROS (Ponnekanti et al., 2001 ; Johanson et al., 2002) développée à l'université de Stanford. Cette plate-forme est construite autour de la gestion d'un tas d'événements (Johanson et al., 2002), c'est-à-dire d'une mémoire non structurée (d'où le terme de tas et non de pile) où les différents événements émis par les participants sont stockés pendant une durée déterminée, et d'où ils peuvent être lus par n'importe quel composant du middleware (ou par un ou plusieurs composants spécifiés). Le tas d'événements iROS se complète naturellement par le gestionnaire de service iCrafter qui correspond à nos besoins en termes d'encapsulation et de déploiement de services (Ponnekanti et al., 2001).

Dans le noyau de notre middleware d'interaction, ce sont ainsi des événements iROS qui sont produits et consommés. Quant aux services qu'il fournit, ce sont des services iCrafter. C'est par la description de ceux-ci que nous commençons l'inventaire des fonctionnalités de notre middleware. Comme tout middleware, un middleware d'interaction se doit de fournir un certain nombre de services indispensables à la gestion d'une architecture et de ses composants. Deux premiers ensembles de services regroupent les services dits « standard » et les services dits « techniques », comme cela apparaît dans la figure 5.

| Services standard | Services techniques |
|---|---|
| Service d'administration | Service pour la déclaration et le déploiement des services des applications (*design-time*) |
| Service de nommage des composants | Service pour l'administration des services d'interaction (*run-time*) |
| Service de gestion de la durée de vie des événements et autres objets internes au middleware | Service de gestion des dispositifs d'interaction (terminal PC, PDA, téléphone) : gestion des connexions et des déploiements |
| Service fournissant des passe-relles pour gérer un déploiement sur plusieurs machines | Services de gestion des données d'interaction : outils pour manipuler, filtrer, transformer des données telles que des énoncés en langage naturel |
| Service de gestion des ressources | Services liés à des outils pour l'interaction, par exemple des analyseurs linguistiques et multimodaux |

Figure 5. Liste des services « standard » et « techniques » de notre middleware

Le troisième et dernier ensemble de services regroupe les services d'interaction. Il s'agit des services qui justifient le concept de middleware d'interaction, et qui matérialisent l'apparition séparée de l'interaction dans MVIC ainsi que le découpage des logiques dont nous avons parlé dans les sections précédentes. La matérialisation du concept HIC dans les systèmes d'interaction et de dialogue homme-machine, c'est la mise en place de services d'interaction qui fédèrent les aspects linguistiques et interactionnels au sein même du middleware, c'est-à-dire à un endroit où les développeurs d'applications n'auront pas à intervenir. Le but est de fournir à ces développeurs un ensemble de services pour les aider à gérer les entrées et les sorties linguistiques ou multimodales de leurs systèmes de dialogue. Un des points essentiels est que, contrairement aux services techniques tels que ceux gérant les données d'interaction, les services d'interaction sont faits pour mettre en œuvre des processus interprétatifs. Autrement dit, les aspects sémantiques du langage naturel sont placés dans les services d'interaction. Désambiguïsation lexicale, résolution des ellipses nominales, résolution des anaphores : autant de processus interprétatifs qui sont incorporés dans les services d'interaction et que les développeurs d'applications pourront paramétrer et exploiter. Si à terme nous visons des services d'interaction incorporant une partie des processus de traitement automatique des langues (Landragin, 2004), nous n'avons implémenté pour l'instant qu'un seul service d'interaction dans le cadre de notre application (à titre de validation du principe) : celui qui constitue le noyau proprement dit du middleware d'interaction, c'est-à-dire son principal point d'entrée. Ce service principal, nommé « IMServ », propose deux méthodes d'appel. La première, « InteractionRequest », sert de point d'entrée pour les événements provenant de l'utilisateur ou de l'un des utilisateurs. Elle est appelée par le conteneur d'interaction chargé de l'interface homme-machine, à chaque fois que cette interface capture une action de l'utilisateur. Que cette action soit un clic souris, un geste sur écran tactile, une commande vocale, ou quoi que ce soit compte tenu des dispositifs disponibles, elle va être codée dans une donnée qui servira de paramètre lors de l'appel à InteractionRequest. La deuxième méthode d'appel au service IMServ, nommée « BusinessRequest », est l'équivalent de la précédente pour une application métier, c'est-à-dire pour un serveur de données ou n'importe quelle application susceptible d'émettre des informations de temps à autre (telles que des notifications de mises à jour, par exemple). Ici aussi, le principal paramètre de l'appel est l'information émise. Avec ces deux méthodes, les comportements des utilisateurs et des applications sont transmis au middleware d'interaction. Celui-ci va alors les traiter, c'est-à-dire vérifier leur nature et les mettre en correspondance avec les possibilités prévues et codées dans les modèles de tâche et dans les profils des utilisateurs. Si l'action est autorisée, un BIP (Business Interaction Pattern) est exécuté, c'est-à-dire que le middleware demande à l'application concernée quelle est l'action-réponse à exécuter face à l'action reçue. Cette action-réponse, par exemple la suppression d'une donnée ou l'exécution d'une commande, est alors lancée. Son résultat est transmis à IMServ, et celui-ci peut notifier le composant à l'origine de l'appel. Une notification est également émise si le BIP ne peut pas être exécuté.

Ce principe permet de dissocier les spécificités des applications des spécificités de l'interaction :
- les applications décrivent au middleware leurs possibilités et leurs contraintes sous la forme de modèles de tâche ;
- les classes d'utilisateurs, les utilisateurs eux-mêmes et leurs préférences sont également décrits dans des profils qui sont fournis au middleware ;
- le code d'une application reste par contre hors middleware (il n'a pas besoin d'y être intégré : seuls comptent les BIP (Business Interaction Patterns) (Faure et al., 2004), c'est-à-dire les primitives permettant de l'exécuter, qui, elles, sont déclarées au middleware) ;
- tous les aspects interactifs sont alors pris en charge par le middleware d'interaction, plus précisément par son noyau, le service IMServ.

Sans entrer dans les détails d'implantation des modèles de tâche – se reporter pour cela à (Faure et al., 2004) – nous notons qu'un modèle de tâche est une sorte de graphe à base d'états et de transitions, chaque transition étant liée à un BIP, issu du formalisme des graphes à états (Harel, 1987). Si par exemple un utilisateur doit se connecter à une application X pour pouvoir l'utiliser, le modèle de tâche de X doit comprendre au moins un état « connecté », un état « déconnecté », et un ensemble de transitions possibles entre les deux, chacune de ces transitions étant liée à un type de donnée caractérisant une action utilisateur. Certaines sont explicitées dans la figure 6. En ce qui concerne le profil d'une classe d'utilisateurs, il s'agit d'un ensemble de données traduisant les droits des membres de cette classe. Le profil d'un utilisateur est de même un ensemble de données, dont son identifiant, sa classe, et une liste de préférences telles que les personnalisations des interfaces homme-machine qu'il utilise, ou encore les personnalisations des données gérées par une application (un utilisateur peut par exemple donner un nom personnalisé à une donnée ou à une action applicative).

```xml
<?xml version="1.0" encoding="ISO-8859-1" ?>
<!-- Fichier de description de la tache pour l'airline station manager   -->
<task_model xmlns:xsi="http://www.w3.org/2001/XMLSchema-instance" xsi:noNamespaceSchemaLocation="task.xsd">
    <starting_state id="disconnected" />
    <state id="disconnected">
        <events>
    </state>
    <state id="connected">
    <state id="browsing_general_templates1">
    <state id="writing_general_msg1">
    <state id="browsing_specific_templates1">
        <events>
            <event id="cancel_specific_msg">
            <event id="select_specific_template">
                <in_param id="message_template" type="business.cofos.data.Template" />
                <interaction_call id="select_specific_template1">
                    <method id="hic.im.business.cofos.bip.common.SelectSpecificTemplate" />
                    <next_states>
                        <positive>
                            <out_param id="message_sent" type="java.lang.String" />
                            <next_state id="connected" />
                        </positive>
                        <negative>
                            <out_param id="incomplete_message" type="java.lang.String" />
                            <next_state id="writing_specific_msg1" />
                        </negative>
                    </next_states>
                </interaction_call>
            </event>
        </events>
    </state>
    <state id="writing_specific_msg1">
    <state id="reading_message1">
```

Figure 6. Un extrait (partiellement développé) du modèle de tâche pour un utilisateur de la classe « airline » (par opposition à la classe « handling » qui correspond à un autre type de profil d'utilisateur, et donc à d'autres rôles). A titre indicatif, ce modèle décrit en XML comporte 624 lignes et 20 000 caractères

En simplifiant légèrement, le rôle principal du service IMServ est ainsi la confrontation d'un comportement avec un modèle de tâche et un profil, de manière à accepter (et faire traiter par l'application concernée) ou à refuser ce comportement. De ce fait, le noyau du middleware d'interaction comporte trois composants, dont les rôles sont décrits ici pour un comportement de l'utilisateur :
- un gestionnaire de tâche chargé de répondre à toute requête sur les possibilités d'une application et sur l'état courant d'un utilisateur dans cette application ;
- un gestionnaire de profils chargé de répondre à toute requête sur les droits et préférences d'un utilisateur ;
- un gestionnaire d'interaction chargé de gérer une boucle d'interaction, c'est-à-dire en gros, la succession suivante : réception d'un comportement de l'utilisateur provenant du conteneur d'interaction ; appel du gestionnaire de profils pour vérifier les droits concernés et retrouver la classe de l'utilisateur ; appel au gestionnaire de tâche pour vérifier que l'utilisateur, compte tenu de sa classe et du modèle de tâche associé, est bien dans un état lui permettant ce comportement ; et, si tout va bien, nouvel appel au gestionnaire de tâche pour l'exécution du BIP concerné et pour le passage au nouvel état ; puis notification au conteneur d'interaction.

Comme nous le voyons avec cette dernière étape, le service principal du middleware renvoie des données de retour. Si c'est le conteneur d'interaction qui est à l'origine de l'appel, c'est le conteneur d'interaction qui est notifié par IMServ. Sans entrer dans les détails de ce composant qui est un peu extérieur au middleware, nous noterons que le conteneur d'interaction gère les interfaces homme-machine, en tant qu'intermédiaire entre l'utilisateur et le middleware d'interaction. Il se comporte comme une application, c'est-à-dire qu'il est déclaré et déployé sous la forme d'un service applicatif (comme nous l'avons vu, l'un des services techniques du middleware permet de procéder à cette déclaration et à ce déploiement). Grâce à ce principe d'encapsulation des échanges, le middleware d'interaction ne fait pas distinction entre ce qui est émis par les utilisateurs ou par les applications. Le but à long terme est que les deux soient interchangeables, c'est-à-dire qu'un utilisateur puisse être remplacé par une application si besoin est, sans que cela ne change quoi que ce soit au niveau du middleware. Dans les systèmes de surveillance ou même de contrôle-commande, l'intérêt est qu'un utilisateur en état de fatigue ou de surcharge cognitive puisse être temporairement remplacé par une application. Cette application constitue alors un mode de fonctionnement minimal (sans doute beaucoup moins performant et pertinent) de la tâche habituelle de l'utilisateur.

Déployé comme un service applicatif, le conteneur d'interaction se doit donc de fournir un certain nombre de méthodes d'appel. Dans notre implantation actuelle, le service « ICServ » fournit une seule méthode d'appel : « DisplayRequest » qui, comme son nom l'indique, fournit un service d'affichage via une ou plusieurs des interfaces homme-machine gérées par le conteneur. Le principal paramètre de cette méthode d'appel est la donnée à afficher. C'est le conteneur d'interaction qui, à partir de cette donnée, se charge de l'adapter au type de terminal. Nous allons arrêter ici la description technique du middleware d'interaction, pour nous étendre un peu sur la manière dont nous l'avons implémenté dans le cadre d'une application métier très spécifique et dans un but clairement défini d'illustration de la séparation entre interface homme-machine et application.

### 5.3. Application

Comme nous l'avons détaillé dans (Sedogbo et al., 2003), notre approche se caractérise avant tout par une séparation claire entre le dialogue homme-machine ou interaction homme-machine d'un côté, avec ses spécificités liées aux modalités de communication et au traitement du langage naturel, et l'application métier de l'autre côté, avec ses contraintes et ses fonctionnalités. Ce n'est qu'en séparant ces deux composants et en clarifiant leur interface qu'on peut espérer d'une part, une réutilisabilité efficace des composants (une même interface multimodale fonctionnant avec plusieurs applications, une même application s'exploitant à l'aide de plusieurs interfaces), et d'autre part, une évolution réaliste vers du dialogue homme-machine multi-applicatif, c'est-à-dire des systèmes capables d'identifier automatiquement l'application adéquate (ou l'application la plus efficace) face à une action de l'utilisateur. C'est en suivant cette approche de séparation que nous avons implémenté le middleware d'interaction, avec l'écriture des services IMServ et ICServ dont nous avons parlé, ainsi que du service « AppliServ » correspondant au point d'entrée de l'application métier.

Cette application relève du domaine de la gestion du trafic aérien. Elle met en œuvre une base de données qui contient les caractéristiques des différents vols et qui est très régulièrement mise à jour, et, dans son état initial, une interface homme-machine présentant ces données sous la forme d'un tableau un peu évolué (les alertes comme les retards sont mis en valeur, par exemple). L'application et son interface telles que nous en avons pris connaissance font partie d'un vaste projet autour du système COFOS (Collaborative Flight Operation System), développé par THALES ATM (Air Traffic Management), et dont le but est de promouvoir les aspects collaboratifs dans l'interaction entre contrôleurs aériens, personnels à l'embarquement (compagnies aériennes) et personnels sur les pistes (aéroport). L'interface initiale est la même pour tout le monde, et l'intérêt de l'étude et de l'implantation que nous avons ensuite effectuée est de montrer qu'il est possible d'intercaler entre la base des données aériennes et cette interface une couche qui permet de gérer de manière transparente et efficace les caractéristiques de l'utilisateur (membre d'une compagnie aérienne ou de l'aéroport) et de son matériel (PC, PDA, ou téléphone portable).

Ainsi, nous avons implémenté notre middleware d'interaction avec un service « COFOSServ », et nous avons démontré son efficacité lorsqu'il s'agit de traduire l'information provenant de la base de données pour l'utilisateur concerné et compte tenu des limitations techniques ou interactives de son matériel. En particulier, un utilisateur sur la piste n'est équipé que d'un téléphone portable sur lequel il n'est pas possible d'afficher un tableau de données très large, alors qu'un utilisateur dans un terminal d'embarquement peut utiliser un PC permettant un affichage bien plus complet.

Par ailleurs, certains utilisateurs peuvent modifier des données (et mettre ainsi à jour la base) alors que d'autres ont juste le droit de prendre connaissance de ces données. Tous ces aspects qui n'étaient pas pris en compte dans le système initial le sont dans le système augmenté du middleware d'interaction. La démonstration est importante, d'une part, parce qu'elle montre qu'il n'a pas été nécessaire de refaire tout le système pour en augmenter la souplesse et les fonctionnalités, et d'autre part, parce qu'elle montre que le middleware d'interaction va dans le sens de la généricité et de la réutilisabilité.

Au cœur du démonstrateur se trouvent trois composants essentiels : le gestionnaire d'interaction, le gestionnaire de tâche et le gestionnaire des profils. Aux extrémités se trouvent d'un côté le serveur de données COFOS, et de l'autre côté les différents terminaux pour la communication homme-machine (terminal PC, PDA, téléphone portable). Comme nous l'avons dit, les rôles principaux du middleware d'interaction sont d'une part, d'adapter les informations provenant de COFOS à un utilisateur et à un terminal particulier, et d'autre part, de contraindre les possibilités de l'utilisateur selon son type (ou classe) et selon une succession d'étapes prédéfinie dans un modèle de tâche. En laissant de côté l'adaptation au terminal qui n'entre pas dans le cadre de ce travail, ces adaptations et ces contraintes requièrent donc deux types de ressources :
– des profils utilisateur spécifiant leurs caractéristiques et préférences ;
– des modèles de tâche, un par classe d'utilisateur.

Le gestionnaire de profils s'occupe des premiers, le gestionnaire de tâche des seconds (c'est lui qui parcourt les modèles de tâche comme celui présenté dans la figure 6, et qui gère les BIP associés), et le gestionnaire d'interaction de leur exploitation lors de l'intervention d'une action de l'utilisateur ou de l'émission d'une information par COFOS. Le gestionnaire d'interaction supervise ainsi la succession des différentes vérifications, adaptations et contraintes. Il repose sur le middleware d'interaction tel que nous l'avons décrit. Tous ces éléments apparaissent dans des fenêtres distinctes sur la figure 8. Nous noterons pour finir que, bien que le système COFOS (dans sa version initiale comme dans notre version) reste un prototype fonctionnel et n'ait pas été utilisé effectivement en situation réelle, il n'en démontre pas moins l'intérêt de notre approche et en constitue une validation. Cette validation est effective au niveau des fonctionnalités du middleware d'interaction, et est renforcée par le fait que nous ne disposions que du code exécutable de l'application COFOS initiale (et non des sources), ce qui nous a obligé à garder une stricte séparation entre les spécificités de l'application métier et les spécificités de l'interaction homme-machine.

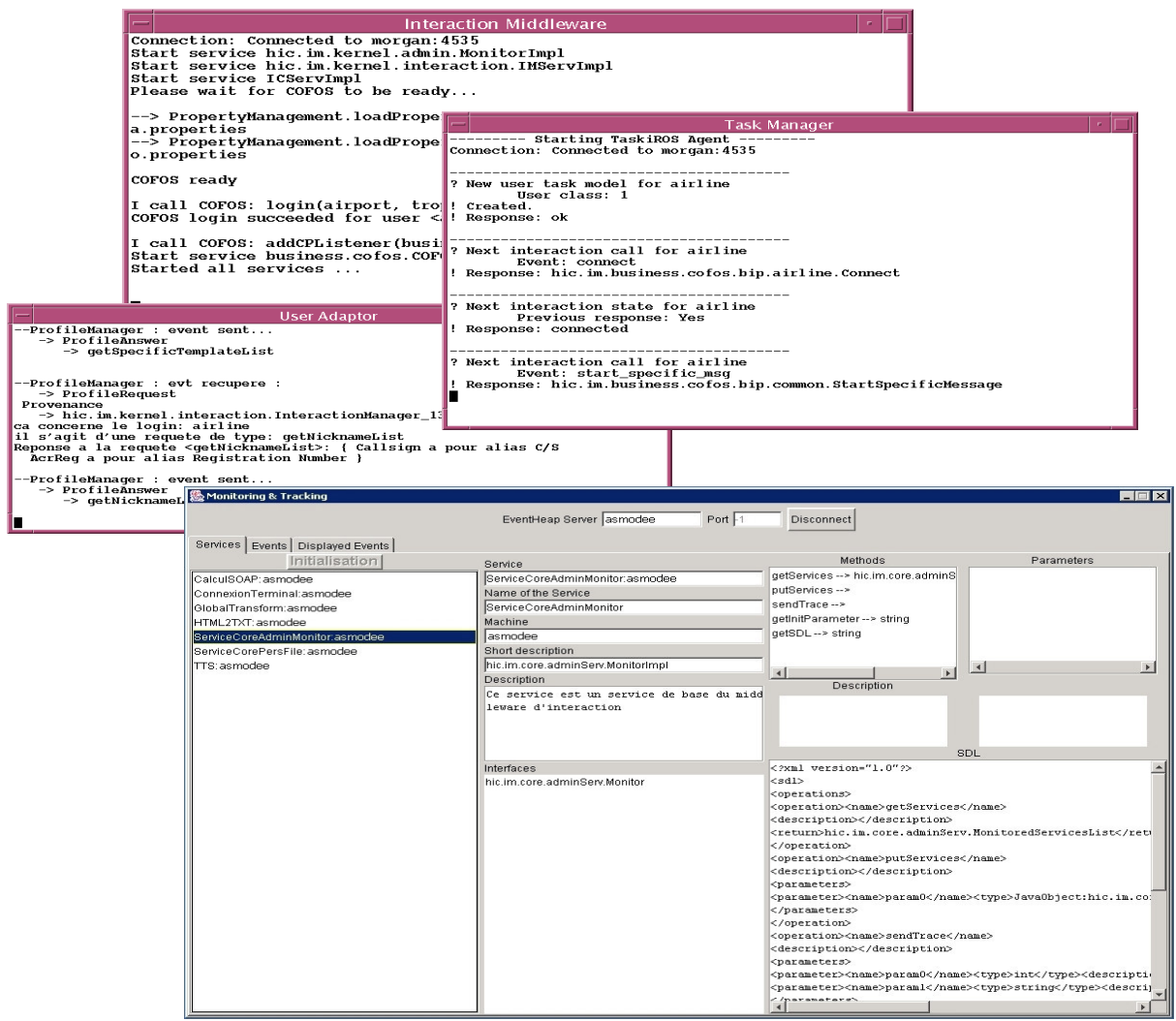

Figure 8. Fenêtres apparaissant lors du lancement du middleware d'interaction (de haut en bas) : fenêtre principale ; gestionnaire de tâche ; gestionnaire de profils ; visualisateur de traces (pour vérifier la liste des services déployés et les appels)

## 6. Conclusion et perspectives

Nous avons proposé un nouveau type d'architecture logicielle pour les systèmes interactifs complexes. Cette architecture est bâtie à partir de modèles de l'état de l'art tels que MVC. L'atelier logiciel reprend ces principes pour les proposer à des équipes de développeurs de prototypes d'interfaces graphiques.

Un nouveau composant intègre la gestion de l'interaction ainsi que sa définition. Nous avons introduit un nouveau tiers médiateur de l'interaction entre les couches de présentation et celles du comportement des applications. Ce tiers médiateur de l'interaction a été intégré dans un composant logiciel dédié de l'atelier JAGUAR© existant, appelé DisplayController. Le DisplayController intègre la logique d'interaction qui représente donc les interactions possibles de l'utilisateur avec le système en dehors de considérations graphiques. Cette interaction est représentative des tâches que l'utilisateur peut ou doit réaliser avec le système informatique.

Ce tiers interaction a également été l'objet d'une implantation sous la forme d'un middleware appelé middleware d'interaction. Ce middleware fournit des services génériques pour l'interaction homme-machine et fournit une abstraction des spécificités des applications d'une part, et des contextes d'utilisation, d'autre part.

Grâce à sa matérialisation sous forme de deux concepts, l'approche HIC peut être perçue comme un cadre général pour l'étude et la conception de l'interaction au-dessus des modèles d'interaction. Nous avons fourni plus de réutilisabilité et plus d'adaptabilité sans intervenir au niveau applicatif. Ces approches permettent de s'abstraire dans une certaine mesure des modalités d'interaction fournies par les applications utilisées, en conséquence de quoi la multimodalité peut être introduite plus facilement comme moyen d'interaction.

Nous souhaitons poursuivre et appliquer cette approche qui propose déjà une vision distribuée de l'interaction au domaine CDM (Collaborative Decision Making), en essayant d'isoler les services génériques utiles pour des applications collaboratives. Nous souhaitons aller d'une approche utilisateur/interaction/système à une approche équipage/interaction/système, en insistant sur l'importance de l'interaction humaine médiatisée par la machine plus que sur l'interaction homme-machine. Par cela nous espérons pouvoir répondre aux questions suivantes : quelles sont les informations que le preneur de décision a besoin de connaître (au niveau local mais aussi au niveau distribué) ? Où l'information est-elle le plus facilement accessible ? Qui peut fournir cette information ? Sous quelle forme peut-on obtenir cette information ?